\documentclass[aps,prd,preprint,superscriptaddress,showpacs,floatfix,nobibnotes]{revtex4-1}

\usepackage{latexsym}
\usepackage{amsmath}
\usepackage{amssymb}
\usepackage{graphicx}
\usepackage{float}
\usepackage{setspace}
\usepackage{longtable}
\usepackage{footnote}
\usepackage{slashed}
\usepackage{diagbox}

\usepackage{bm}
\usepackage{epsfig}
\usepackage{subfigure}
\newcommand{\bea}{\begin{eqnarray}}
\newcommand{\eea}{\end{eqnarray}}

\begin{document}

\title{The effect of a Chern-Simons term on dynamical gap generation in graphene}

\author{M.E. Carrington}
\email[]{carrington@brandonu.ca} \affiliation{Department of Physics, Brandon University, Brandon, Manitoba, R7A 6A9 Canada}\affiliation{Winnipeg Institute for Theoretical Physics, Winnipeg, Manitoba}

\date{January 11, 2019}

\begin{abstract}
  We study the effect of a Chern-Simons term on dynamical gap generation in a low energy effective theory that describes some features of mono-layer suspended graphene. 
  We use a non-perturbative Schwinger-Dyson approach. 
  We solve a set of coupled integral equations for eight independent dressing functions that describe fermion and photon degrees of freedom. 
  We find a strong suppression of the gap, and corresponding increase in the critical coupling, as a function of increasing Chern-Simons coefficient. 
\end{abstract}


\normalsize
\maketitle

\section{Introduction}

Quantum electrodynamics in 2+1 dimensions (QED$_{2+1}$) has been studied for many years as a toy model for quantum chromodynamics (QCD). 
The main point is that QED$_{2+1}$ is strongly coupled and therefore, in spite of being abelian, it can be used to study many interesting features of QCD \cite{appelquist1,pisarski,fradkin,gusynin,teber}. 
%
In this paper, we are interested in reduced QED$_{3+1}$ (RQED) in which the fermions are restricted to remain in a two dimensional plane, but the photons which are responsible for the interactions between fermions, are not. 
In the reduced theory the coulomb interaction between the electrons has the same $1/r$ form as in the (3+1) dimensional theory, instead of the logarithmic form obtained from QED$_{2+1}$. 
The theory is physically relevant for the description of what are called Dirac planar materials, which refer to condensed matter systems for which the underlying lattice structure produces a fermionic dispersion relation that has the form of a Dirac equation, in some regimes. 
We are particularly interested in graphene, where the fermions have an
effective speed $v_F$ which is on the order of 300 times smaller than the speed of light. 
The unique  band structure of graphene gives it high mobility, large thermal and electrical conductivity, and optical transparence, which are characteristics that are valuable in technological applications. 
We study specifically suspended single layer graphene, where we deal with a single atomic layer in the absence of scattering from a substrate, so that the intrinsic electronic properties of the system are accessed. For simplicity we will also work at half filling (which means zero chemical potential).

In both QED$_{2+1}$ and RQED the fermions couple to  a three dimensional abelian gauge field, and therefore the Chern-Simons (CS) term can be added to the action. This term breaks the time reversal symmetry, and gives a mass to the photon. It is important in condensed matter physics in the context of chiral symmetry breaking \cite{appelquist2,matsuyama,bashir,kondo}, high temperature superconductivity \cite{highTc}  and the Hall effect \cite{Witten-review}.
CS terms can dynamically generate magnetic fields in QED$_{2+1}$ \cite{hosotani}, and magnetic fields are thought to influence dynamical symmetry breaking in a universal and model independent way through what is known as magnetic catalysis (for a review see \cite{shovkovy}). 
In this work we have used RQED and studied the importance of a CS term on phase transitions in graphene. 

The coupling constant and CS parameter are dimensionful scales in QED$_{2+1}$, but they are  dimensionless parameters in RQED. 
In natural units the effective coupling can be written  $\alpha = e^2/(4\pi\epsilon v_F)$ where $v_F \sim c/300$ is the velocity of a
massless electron in graphene. The parameter $\epsilon\ge 1$ is related to the screening properties of the graphene
sheet and we take the vacuum value $\epsilon=1$. The Chern-Simons parameter will be denoted $\theta$, and we consider $\theta\in(0,1)$.

\section{The low energy effective theory}

\subsection{Non-interacting Hamiltonian}

The carbon atoms in graphene are arranged in a 2-dimensional hexagonal lattice. 
The hexagonal structure can be viewed as two sets of interwoven triangular sublattices (called $A$ and $B$). The geometry dictates each primitive cell has one atom from the $A$ sublattice and one from the $B$ sublattice, and that each lattice site has three nearest neighbours on the opposite sublattice. 
For each atom, three of the four outer electrons form hybridized $\sigma$-bonds with the three nearest neigbours. The fourth sits in the $p_z$ orbital, perpendicular to the hybrid orbitals, and forms a $\pi$-bond. The simplest description of graphene is a tight binding Hamiltonian for the $\pi$-orbitals
\bea
\label{tight}
H_0 = -t\sum_{\langle \vec n \vec n^\prime \rangle \sigma}
\big[ a_{\vec n\sigma}^\dagger b_{\vec n^\prime \sigma} + {\rm h.c} \big]\,
\eea
 where $t$ is the nearest neighbour hopping parameter and the operators $a_{\vec n \sigma}^\dagger$ and $b_{\vec n' \sigma}^\dagger$ are creation operators for $\pi$ electrons with spin $\sigma$ on the $A$ and $B$ sublattices, respectively.

We can rewrite the Hamiltonian as a momentum integral by Fourier transforming.
Our definitions for the lattice vectors and discrete Fourier transforms are given in Appendix \ref{notation}.
From the dispersion relation for the non-interacting theory we obtain six $K$ points, and our choice of two inequivalent ones, which we denote $K_\pm$,  is given in equation (\ref{Kpoints}). 
Using equations (\ref{fourier1}, \ref{fourier2}) we rewrite the Hamiltonian in (\ref{tight}) as a momentum integral and  we expand around $K_\pm$. 
We define a 4 component spinor: 
\bea
\label{spinor}
\Psi_\sigma(\vec p) = (a_\sigma(\vec K_+ + \vec p),b_\sigma(\vec K_+ +\vec p),b_\sigma(\vec K_- +\vec p),a_\sigma(\vec K_- +\vec p))^T
\eea
where the superscript $T$ indicates that the spinor should be written as a column vector. 
Using this notation the tight binding Hamiltonian becomes
\bea
\label{tight-momentum}
H_0 = \hbar v_F\sum_\sigma \int\frac{d^2 p}{(2\pi)^2} \bar\Psi_\sigma(\vec p) (\gamma^1 p_1 + \gamma^2 p_2)\Psi_\sigma(\vec p)
\eea
where we have defined $\hbar v_F = 3at/2$.
The Lagrangian of the effective theory (including minimal coupling to the gauge field) then takes the form \cite{carbotte}
\bea
\label{L0}
{\cal L} = \sum_\sigma \bar\Psi_\sigma(t,\vec x)\big[i\gamma^0 D_t + i\hbar v_F \vec\gamma\cdot\vec D\big]\Psi_\sigma(t,\vec x)
\eea
where we define $D_\mu = \partial_\mu - i e A_\mu$ (taking $e>0$).

In the next sections we will discuss how to include interactions. 
At this point however, we note that while our  effective theory can accurately describe the low energy dynamics of the system and allow us to correctly include both frequency and non-perturbative effects, it does
not allow for the inclusion of screening from the $\sigma$-band electrons and localized higher energy states.

\subsection{Symmetries}
We consider the discrete symmetries of the tight binding Hamiltonian. 
The parity, time reversal and charge conjugation transformations on the spinor in (\ref{spinor}) are
\bea
\label{parity-def}
&& P\Psi(\vec p) P^{-1} = \gamma^0 \Psi(-\vec p) \\
\label{time-rev-def}
&& T\Psi(\vec p) T^{-1} = i \sigma_2 \gamma^1 \gamma^5  \Psi(-\vec p) \\
\label{charge-c-def}
&& C\Psi(\vec p) C^{-1} = \gamma^1 \bar\Psi (\vec p)^T\,.
\eea
It is easy to check that the non-interacting theory is invariant under these symmetries. 
To see the physical content of equations (\ref{parity-def} - \ref{charge-c-def})  we show the action of each on the spinor defined in equation (\ref{spinor}). 

The parity transformation takes the form
\bea
\Psi = 
\left(\begin{array}{c} a_\sigma(K_+ + \vec p) \\b_\sigma(K_+ + \vec p) \\b_\sigma(K_- + \vec p) \\a_\sigma(K_- + \vec p) \end{array}\right)
~\overset{\displaystyle P}{\longrightarrow}~
\left(\begin{array}{c} b_\sigma(K_- - \vec p) \\a_\sigma(K_- - \vec p) \\a_\sigma(K_+ - \vec p) \\b_\sigma(K_+ - \vec p) \end{array}\right)
\eea
which tells us that the parity operator reverses the sign of the momentum and exchanges the sublattices. 
We note that this definition is different from the one commonly used in  QED$_{2+1}$, where the transformation ${\cal P}:(x,y) \to (-x,-y)$ would correspond to spatial rotation. Because of the hexagonal lattice structure of graphene, spatial rotation is not a symmetry of the system unless the sublattice indices are interchanged. 

The time reversal operator changes the sign of momentum and spin and its action on a spinor is 
\bea
\Psi = 
\left(\begin{array}{c} a_\sigma(K_+ + \vec p) \\b_\sigma(K_+ + \vec p) \\b_\sigma(K_- + \vec p) \\a_\sigma(K_- + \vec p) \end{array}\right)
~\overset{\displaystyle T}{\longrightarrow}~
\left(\begin{array}{c} a_\sigma(K_- - \vec p) \\b_\sigma(K_- - \vec p) \\b_\sigma(K_+ - \vec p) \\a_\sigma(K_+ - \vec p) \end{array}\right)
\eea
(where we have not explicitly written the action of the factor $i\sigma_2$ which flips spin) and therefore the time reversal operator inverts the $K$ points (and spin) but does not act on the sublattice degrees of freedom. 

The action of the charge conjugation operator is 
\bea
\Psi = 
\left(\begin{array}{c} a_\sigma(K_+ + \vec p) \\b_\sigma(K_+ + \vec p) \\b_\sigma(K_- + \vec p) \\a_\sigma(K_- + \vec p) \end{array}\right)
~\overset{\displaystyle C}{\longrightarrow}~
\left(\begin{array}{c} -b^\dagger_\sigma(K_+ + \vec p) \\-a^\dagger_\sigma(K_+ + \vec p) \\a^\dagger_\sigma(K_- + \vec p) \\b^\dagger_\sigma(K_- + \vec p) \end{array}\right)\,.
\eea

We can also consider continous symmetries of the low energy effective theory. 
The action is invariant under the enlarged group of global symmetries generated by both $\gamma_5$ and the third spatial gamma matrix ($\gamma_3$) which is not part of the Lagrangian (\ref{L0}). 
The matrices 
\bea
T_1 = \frac{i}{2} \gamma^3 \,, ~~
T_2 = \frac{1}{2} \gamma^5 \,, ~~
T_3 = \frac{i}{2} \gamma^3\gamma^5 
\eea
commute with the Hamiltonian. They also  
satisfy the commutation relations $[T^i, T^j] = i \epsilon^{ijk}T^k$ and therefore form a four dimensional representation of $SU(2)$. Including $T_4 = \mathbb{I}/2$ gives a representation of $U(2)$. Physically this is a symmetry in the space of valley and sublattice indices, where `valley' refers to the $K_\pm$ points.
The non-interacting theory has a global $U(4)$ symmetry that operates in the space of [valley $\otimes$ sublattice $\otimes$ spin].
We call this a chiral symmetry and, using our representation of the gamma matrices (see Appendix \ref{notation}), the chirality quantum number corresponds to the valley index. 

\subsection{Fermion bilinears}
\label{bilinears}



%
One reason that fermion bilinears are interesting is that, close to the critical point, possible interactions of the low energy theory are constrained to have the form of local four-fermion interactions. 
For example, in the Gross-Neveu model the basic interaction is a four-fermi contact between scalar or
pseudoscalar densities, and in the Thirring model the interaction is a contact between two conserved currents. We note that while short range interactions are not relevant for dynamics in a perturbative theory,  they can be important in a strongly coupled system. 
Mass scales are especially interesting because they are directly related to chiral symmetry breaking and a possible semi-metal/insulator transition. 

We use $\Gamma^{(n)}$ to indicate one element of the list
\bea
\label{gammabasis}
\Gamma = \{\mathbb{I},\gamma^\mu,\gamma^3,i\gamma^5,i\gamma^\mu\gamma^\nu,i\gamma^\mu\gamma^3,\gamma^\mu\gamma^5,\gamma^3\gamma^5\}
\eea
with $\mu \in (0,1,2)$, which gives a complete basis in Dirac space.
We define a set of fermion bilinears as
\bea
\label{bilinear-eff}
{\cal G}^{(n)} = m^{(n)}\int d^2 x \; \bar\Psi(\vec x) \,\Gamma^{(n)} \,\Psi(\vec x)\,.
\eea
The terms 
constructed with scalar/pseudo-scalar elements $\Gamma^{(n)} \in \{ 1,\gamma^3,i\gamma^5,\gamma^3\gamma^5\}$ correspond to mass terms and will be denoted ${\cal M}$, ${\cal M}^3$, ${\cal M}^5$ and ${\cal M}^{35}$.
%

We look at the transformation properties of  fermion bilinears under parity, time reversal and charge conjugation. 
We introduce the notation $\gamma^{\tilde\mu} = (\gamma^0,-\gamma^i)$ with $i\in(1,2)$. Two examples where this notation can be used are: $P \bar\Psi \gamma^\mu \Psi  P^{-1} = \bar\Psi \gamma^{\tilde \mu} \Psi $ and $P   \bar\Psi \gamma^\mu\gamma^5 \Psi  P^{-1} = -  \bar\Psi \gamma^{\tilde \mu}\gamma^5 \Psi$. 
Our results are shown in table \ref{table-PCT}, and transformations of the type discussed above are listed with a tilde over the sign. 
The first of the two examples given above is written $\tilde{+}$ in the second row of the first column of table \ref{table-PCT}, and the second is the symbol $\tilde -$ in the seventh row of the first column. 
\begin{table}[h]
\begin{center}
\begin{tabular}{|c|c|c|c|}
\hline 
\vspace*{-.2cm} & & & \\ 
~~~~~~~~~~~  & ~~~ $P$ ~~~ &  ~~~ $C$ ~~~ & ~~~ $T$ ~~~ \\
\vspace*{-.2cm} & & & \\ 
 \hline
 \vspace*{-.2cm} & & & \\ 
 $\mathbb{I}$ & + & + & +  \\
 \vspace*{-.2cm} & & & \\ 
 \hline 
 \vspace*{-.2cm} & & & \\ 
 $\gamma^\mu$ & $\tilde +$ & - & $\tilde +$    \\
 \vspace*{-.2cm} & & & \\ 
 \hline 
 \vspace*{-.2cm} & & & \\ 

 $\gamma^3$ & - & + & + \\
 \vspace*{-.2cm} & & & \\ 
 \hline 
 \vspace*{-.2cm} & & & \\ 

 $i \gamma^5$ & - & - & +   \\
 \vspace*{-.2cm} & & & \\ 
 \hline 
 \vspace*{-.2cm} & & & \\ 

 $i\gamma^\mu\gamma^\nu$ & $\tilde +$ & - & $\tilde -$    \\
 \vspace*{-.2cm} & & & \\ 
 \hline 
 \vspace*{-.2cm} & & & \\ 

 $i\gamma^\mu\gamma^3$ & $\tilde -$ & + & $\tilde -$   \\
 \vspace*{-.2cm} & & & \\ 
 \hline 
 \vspace*{-.2cm} & & & \\ 

 $\gamma^\mu\gamma^5$ & $\tilde -$ & - & $\tilde -$  \\
 \vspace*{-.2cm} & & & \\ 
 \hline 
 \vspace*{-.2cm} & & & \\ 

 $\gamma^3\gamma^5$ & + & + & -   \\
 \vspace*{-.01cm} & & & \\ 
 \hline 
 \end{tabular}
\end{center}
\caption{Transformation properties of the bilinears defined in equation (\ref{bilinear-eff}) under $P$, $C$, $T$. \label{table-PCT}}
\end{table}
%
%
The mass terms ${\cal M}^3$ and ${\cal M}^5$ can be accessed from the standard Dirac mass ${\cal M}$ by a change of integration variables in the path integral. 
The mass ${\cal M}^{35}$ is completely independent of the other three, and is related to a model introduced by Haldane \cite{haldane86}.
We remark that although actions constructed from an effective Lagrangian with mass term ${\cal M}$, ${\cal M}^3$ or ${\cal M}^5$ will describe identical physics, the symmetries of a continuous theory are not necessarily evident in the original discrete theory, which means that equivalent continuous theories may correspond to different discrete theories.

To see directly how mass terms are related to physical quantities in the discrete theory, we look at a specific example. We consider a term in the Hamiltonian of the form
\bea
\label{H1-dis}
H_1 = \sum_{\vec n \sigma}\big[m_a a^\dagger_{\vec n\sigma}a_{\vec n\sigma} + m_b b^\dagger_{\vec n\sigma}b_{\vec n\sigma} \big]\,,
\eea
which would correspond to different densities of particles on the $A$ and $B$ sublattices,  and could be realized physically by placing the graphene sheet on a substrate. 
Fourier transforming to momentum space and expanding around the $K$ points, equation (\ref{H1-dis}) becomes 
\bea
\label{HAVEmass}
H_1 =  \sum_\sigma \int \frac{d^2p}{(2\pi)^2}\,\big(m_+ \bar\Psi_\sigma(\vec p) \gamma_0 \Psi_\sigma(\vec p) + m_- \bar\Psi_\sigma(\vec p) \gamma_3 \Psi_\sigma(\vec p) \big)\,,
\eea
where we have defined $m_\pm = \frac{1}{2}(m_a\pm m_b)$.  
The term in (\ref{HAVEmass}) with the factor $m_-$ is proportional to the ${\cal M}^3$ mass term, which breaks parity. Writing it explicitly in terms of creation and annihilation operators we obtain
\bea
{\cal M}^{3} = && \int d^2x\,\bar\Psi \gamma^3\Psi \nonumber\\
\label{M3expansion}
= && \sum_\sigma \int\frac{d^2p}{(2\pi)^2}\,\bigg(\big[a_\sigma^\dagger(\vec K_+ +\vec p)a_\sigma(\vec K_+ +\vec p) + a_\sigma^\dagger(\vec K_- + \vec p)a_\sigma(\vec K_- + \vec p)\big] \\
&& ~~~~~~~~~~~~~~ -  \big[(b_\sigma^\dagger(\vec K_+ +\vec p)b_\sigma(\vec K_+ +\vec p) + b_\sigma^\dagger(\vec K_- + \vec p)b_\sigma(\vec K_- + \vec p)\big] \bigg)\nonumber
 \eea
which makes clear that the order parameter ${\cal M}^3$ is proportional to the difference in electron densities for the $A$ and $B$ sublattices. A non-zero value of this order parameter corresponds physically to a charge density wave, and lifts the sublattice degeneracy. 
The term in (\ref{HAVEmass}) that is proportional to $m_+$ is less interesting, since it can be absorbed into a redefinition of the chemical potential.

The independent mass term
\bea
{\cal M}^{35} = && \int d^2 x\;\bar\Psi \gamma^3\gamma^5 \Psi \nonumber\\
\label{M8expansion}
= && \sum_\sigma \int\frac{d^2p}{(2\pi)^2}\,\bigg(\big[a_\sigma^\dagger(\vec K_+ +\vec p)a_\sigma(\vec K_+ +\vec p) - a_\sigma^\dagger(\vec K_- + \vec p)a_\sigma(\vec K_- + \vec p)\big] \\
&& ~~~~~~~~~~~~~~ - \big[(b_\sigma^\dagger(\vec K_+ +\vec p)b_\sigma(\vec K_+ +\vec p) - b_\sigma^\dagger(\vec K_- + \vec p)b_\sigma(\vec K_- + \vec p)\big] \bigg)\,\nonumber
\eea
corresponds to a gap with opposite sign at the $K_-$ point, relative to 
${\cal M}^{3}$. 
Mathematically a triangular next-neighbour hopping term in the Hamiltonian of the discrete theory gives a mass term proportional to ${\cal M}_{35}$ in the effective theory. This is shown in Appendix \ref{haldane-mass}.
Physically it corresponds to a topologically nontrivial phase generated by currents propagating on the two different sublattices. 
Both the CS term and the ${\cal M}_{35}$ mass term violate time reversal invariance (see Table \ref{table-PCT}) and one therefore expects that one loop radiative corrections to the photon polarization tensor obtained from internal fermions with a Haldane type mass would generate a $T$ odd piece in the polarization tensor, or that including a CS term in the photon part of the action would dynamically generate a Haldane type mass for the fermions. In this paper we 
will introduce a CS term into the action, and study the effect of this term, through dynamical mass generation, on phase transitions in graphene. 

\subsection{The brane action}

Dynamical photons are included in RQED by constructing the brane action \cite{marino,Miransky2001}. We start with the four dimensional Euclidean action
\bea
S = \int d^4 x\big[\frac{1}{4}F_{\mu\nu}F_{\mu\nu} - \frac{1}{2\xi} (\partial_\mu A_\mu)^2+i e \bar\Psi \slashed{A} \Psi\big]
\eea  
and integrate out the four dimensional gauge field to obtain
\bea
&& S\to~~ \frac{1}{2}\int d^4 x\,\int d^4 y\,J_\mu(x) D_{\mu\nu}(x-y) J_\nu(y) \nonumber \\
\label{D4d}
&& D_{\mu\nu}(x-y) = \int\frac{d^3K}{(2\pi)^3} \int \frac{dk_3}{2\pi}\,e^{ik(x-y)}
\left[\delta_{\mu\nu}-(1-\xi)\frac{k_\mu k_\nu}{K^2+k_3^2}\right]\frac{1}{K^2+k_3^2}\,
\eea
where we write $k = (K,k_3)$. We  use capital letters for three vectors which include a time-like component, for example, $K=(k_0,\vec k) =(k_0,k_1,k_2)$ and $X=(x_0,\vec x) = (x_0,x_1,x_2)$.
To describe graphene we take 
\bea
J_3=0  \text{~~and~~} J_\mu(x_0,x_1,x_2,x_3) = j_\mu(x_0,x_1,x_2)\, \delta(x_3)  \text{~~for~~} \mu\in(0,1,2)
\eea
which allows us to do the $k_3$ integral in (\ref{D4d}) analytically and obtain
\bea
\label{D3d}
D_{\mu\nu}(X-Y) = \int\frac{d^3K}{(2\pi)^3} \,e^{iK(X-Y)}
\left[\frac{\delta_{\mu\nu}}{2\sqrt{K^2}}-(1-\xi)\frac{K_\mu K_\nu}{4\sqrt{K^2}K^2}\right]\,.
\eea
Note that in this equation the indices $\mu$ and $\nu$ are $\in(0,1,2)$ and therefore they should properly be written differently (as $\bar\mu$ and $\bar\nu$, for example), but to simplify the notation we use the same letters for these indices. 
We can rescale the gauge parameter $(1-\xi)\to 2(1-\bar\xi)$ and suppress the bar, to remove the factor 1/4 in the last term in (\ref{D3d}).

We introduce a three dimensional vector field (which we again call $A$) and write the effective action
\bea
\label{actionE}
S = \int d^3X \big[\frac{1}{2} F_{\mu\nu}\frac{1}{\sqrt{-\partial^2}}F_{\mu\nu} + A_\mu J_\mu +\frac{1}{\xi} \partial\cdot A \frac{1}{\sqrt{-\partial^2}}\partial\cdot A \big]
\eea
which corresponds to (\ref{D3d}) in the sense that if we integrate out the gauge field we reproduce the dimensionally reduced propagator. 
We redefine the gauge fixing term to be $(\partial\cdot A)^2/\xi$, add the kinetic term for the fermions [see equation (\ref{L0})], and add a CS term to obtain:
\bea
\label{actionE2}
S = \int d^3X \big[\bar\Psi i \slashed{D}\Psi + \frac{1}{2} F_{\mu\nu}\frac{1}{\sqrt{-\partial^2}}F_{\mu\nu} + \frac{1}{2\xi} (\partial\cdot A)^2 + i \theta \epsilon_{\mu\nu\lambda}A_\mu \partial_\nu A_\lambda \big]\,.
\eea

We want to use this relativistic theory to describe graphene near the Dirac points. 
To do this, we replace the Euclidean metric in the first term of equation (\ref{actionE2}) with the non-covariant form
\bea
g_{\mu\nu}~~\to~~ M_{\mu\nu} ~~\text{with}~~
\label{Mdef}
M = 
\left[\begin{array}{ccc}
~1~ & ~0~ & ~0~ \\
0 & v_F  & 0 \\
0 & 0 & v_F   \\
\end{array}
\right]\,
\eea
so that $\bar\Psi i \slashed{D}\Psi$ becomes $\bar\Psi i \gamma_\mu M_{\mu\nu} D_\nu \Psi$. 
We obtain the Feynman rules (in Landau gauge) from the resulting action:
\bea
\label{bareFR}
&& S^{(0)}(p_0,\vec p) = -\big(i\gamma_\mu M_{\mu\nu} P_\nu\big)^{-1}\,,\\[2mm]
&& G^{(0)}_{\mu\nu}(p_0,\vec p)=\bigg(\delta_{\mu\nu}-\frac{P_\mu P_\nu}{P^2}\bigg)\,\frac{1}{2\sqrt{P^2}}\,, \\[1mm]
\label{barevert}
&& \Gamma^{(0)}_\mu = M_{\mu\nu}\gamma_\nu\,.
\eea

From this point on we will not refer again to the original four dimensional theory. We define new notation so that small letters denote  the spatial components of three vectors [for example $P = (p_0,\vec p)$].
We also introduce some additional notational simplifications that will be used in the rest of this paper: we will sometimes write all momentum arguments of functions with a single letter
 [for example
$S(P):=S(p_0,\vec p)$],
 we define $dK :=  dk_0 d^2k /(2\pi)^3$, and we write $Q=K-P$.

\section{Non perturbative theory}

We will include non-perturbative effects by introducing fermion and photon dressing functions, and solving a set of coupled Schwinger-Dyson (SD) equations.
\subsection{Propagators and vertices}

In the non-perturbative theory the bare propagator $S^{(0)}(P)$ in equation (\ref{bareFR}) is written with six dressing functions $(Z^+_P,A^+_P,B^+_P,Z^-_P,A^-_P,B^-_P)$, where we have used subscripts instead of brackets to indicate the  momentum dependence [for example, $Z^+_P:=Z^+(p_0,\vec p)$]. We define two projection operators $\chi_\pm = \frac{1}{2}(1\pm \gamma_3 \gamma_5)$. Using this notation the fermion propagator has the form
\bea
&& S^{-1}(P) = \big[- i (Z^+_P p_0\gamma_0\chi_+   + v_F A^+_P \vec p\cdot \vec \gamma)
+ B_P^+ \big] \chi_+   +
\big[- i (Z^-_P p_0\gamma_0\chi_-
+ v_F A^-_P \vec p\cdot \vec \gamma)
+ B_P^-\big] \chi_-  \nonumber \\[2mm]
&& S(P) = 
\frac{1}{{\rm Den}_P^+} \big[i(Z^+_P p_0\gamma_0  +  v_F A^+_P \vec p\cdot \vec \gamma)  + B_P^+ \big]\chi_+      
+ \frac{1}{{\rm Den}_P^-} \big[i(Z^-_P p_0\gamma_0 +  v_F A^-_P \vec p\cdot \vec \gamma) + B_P^-\big] \chi_-       \nonumber\\[2mm]
&& {\rm Den}_P^{\pm } = p_0^2 Z_P^{\pm 2} + p^2 v_F^2 A_P^{\pm 2} + B_P^{\pm 2}\,.
\label{dressedS}
\eea
We define the even and odd functions:
\bea
X_{\pm} = X_{\rm even} \pm X_{\rm odd} ~~\to~~ X_{{\rm even/odd}} = \frac{1}{2} (X_+ \pm X_-)
\eea
where $X\in(Z,A,B)$.
In the notation of section \ref{bilinears}, 
$B_{\rm even}(0,0)$ is a standard Dirac type mass (denoted ${\cal M}$) which breaks chiral symmetry but not  time reversal symmetry, and $B_{\rm odd}(0,0)$ is a Haldane type mass (${\cal M}^{35}$) which preserves chiral symmetry but violates time reversal invariance. 
In the bare theory $Z_\pm = A_\pm = 1$ and $B_\pm=0$, and therefore the odd functions are zero. It is easy to see that (\ref{dressedS}) reduces to  (\ref{bareFR}) in this limit. 

The feynman rule for the dressed vertex is 
\bea
\label{dressedVert}
 \Gamma_\nu(P,K) = \frac{1}{4}\big[H^+_{\nu\sigma}(P) + H^+_{\nu\sigma}(K)\big]\gamma_\sigma(1+\gamma_5) 
+ \frac{1}{4}\big[H^-_{\nu\sigma}(P) + H^-_{\nu\sigma}(K)\big]\gamma_\sigma(1-\gamma_5) 
\eea
where $P$ is the outgoing fermion momentum, $K$ is the incoming fermion momentum, and $H^\pm$ indicates the diagonal 3$\times$3 matrix
\bea
\label{Amatrix}
H^\pm(P) = 
\left[\begin{array}{ccc}
Z^\pm(P) & 0 & 0 \\
0 & v_F A^\pm(P)  & 0 \\
0 & 0 & v_F A^\pm(P)   
\end{array}
\right]\,.
\eea
It is clear that (\ref{dressedVert}, \ref{Amatrix}) reduce to (\ref{barevert}) in the limit $Z_\pm = A_\pm = 1$. Equations (\ref{dressedVert}, \ref{Amatrix}) are the first term in the full Ball-Chiu vertex \cite{ball-chiu}.
We include only the first term because calculations are much easier using this simpler ansatz, and because in our previous calculation we found that the contribution
of the additional terms is very small \cite{Carrington0}.

To define the dressed photon propagator we start with a complete set of 11 independent projection operators. Defining $n_\mu = \delta_{\mu 0}-q_0 Q_\mu /Q^2$ we write
\bea
&& P^1_{\mu\nu} = \delta_{\mu\nu}-\frac{Q_\mu Q_\nu}{Q^2}\,,~~
P^2_{\mu\nu} = \frac{Q_\mu Q_\nu}{Q^2}\,,~~
P^3_{\mu\nu} = \frac{n_\mu n_\nu}{n^2}\,,~~
P^4_{\mu\nu} = Q_\mu n_\nu \,,~~
P^5_{\mu\nu} = n_\mu Q_\nu \,,~~ \nonumber\\
&& P^6_{\mu\nu} = \epsilon_{\mu\nu\alpha}Q_\alpha\,,~~
P^7_{\mu\nu} = \epsilon_{\mu\nu\alpha}n_\alpha \frac{Q^2}{q^2}\,,~~
P^8_{\mu\nu} = \epsilon_{\mu\alpha\beta}Q_\alpha n_\beta Q_\nu\,,~~
P^9_{\mu\nu} = \epsilon_{\nu\alpha\beta}Q_\alpha n_\beta Q_\mu\,,~~ \nonumber\\
&& P^{10}_{\mu\nu} = -\epsilon_{\mu\alpha\beta}Q_\alpha n_\beta n_\nu \frac{Q^2}{q^2}\,,~~
P^{11}_{\mu\nu} = -\epsilon_{\nu\alpha\beta}Q_\alpha n_\beta n_\mu \frac{Q^2}{q^2}\,.
\eea
Using this notation the inverse dressed photon propagator can be written
\bea
G_{\mu\nu}^{-1} = 2\sqrt{Q^2}\big[P^1_{\mu\nu}+\frac{1}{\xi}P^2_{\mu\nu}\big] + 2\theta P^6_{\mu\nu} + \Pi_{\mu\nu}
\eea
where the polarization tensor is written in a completely general way as the sum 
\bea
\Pi_{\mu\nu} = \sum_{i=1}^{11} a_i P^i_{\mu\nu}\,.
\eea

We invert the inverse propagator and then impose the constraints that the polarization tensor be transverse and satisfy the symmetry condition $\Pi_{\mu\nu}(Q) = \Pi_{\nu\mu}(-Q)$.
The surviving components of the polarization tensor give
\bea
\Pi_{\mu\nu}(Q) = \alpha(Q) P^1_{\mu\nu}+\gamma(Q) P^3_{\mu\nu} + \Theta(Q) P^6_{\mu\nu} + \rho(Q)\,\big[P^{10}_{\mu\nu} + P^{11}_{\mu\nu}\big]
\eea
and the propagator is
\bea
\label{prop}
&& G_{\mu\nu} = G_L P^3_{\mu\nu} + G_T \big[P^1_{\mu\nu} - P^3_{\mu\nu}\big] + G_D P^6_{\mu\nu} + G_E \big[P^{10}_{\mu\nu} - P^{11}_{\mu\nu}\big]\,\\
&& G_L = \frac{2\sqrt{Q^2} + \alpha}
{(2\sqrt{Q^2} + \alpha)(2\sqrt{Q^2} + \alpha+\gamma) + Q^2(2\theta+\rho+\Theta)^2} \nonumber\\
&& G_T = \frac{2\sqrt{Q^2} + \alpha + \gamma}
{(2\sqrt{Q^2} + \alpha)(2\sqrt{Q^2} + \alpha+\gamma) + Q^2(2\theta+\rho+\Theta)^2} \nonumber\\
&& G_D = -\frac{(2\theta+\Theta)(2\sqrt{Q^2} + \alpha + \gamma)}
{(2\sqrt{Q^2} + \alpha)\big[(2\sqrt{Q^2} + \alpha)(2\sqrt{Q^2} + \alpha+\gamma) + Q^2(2\theta+\rho+\Theta)^2\big]} \nonumber \\
&& G_E = \frac{(2\theta+\Theta)(2\sqrt{Q^2} + \alpha + \gamma)-(2\sqrt{Q^2}+\alpha)(2\theta+\rho+\Theta)}
{(2\sqrt{Q^2} + \alpha)\big[(2\sqrt{Q^2} + \alpha)(2\sqrt{Q^2} + \alpha+\gamma) + Q^2(2\theta+\rho+\Theta)^2\big]} \,.\nonumber
\eea

\subsection{Fermion Schwinger-Dyson equations}
\label{ferm-sd-eqns}

The inverse fermion propagator is written generically as
\bea
\label{SF2}
&& S^{-1}(P) = (S^{(0)})^{-1}(P)+\Sigma(P)
\eea 
where the fermion self energy is obtained from the SD equation as
\bea
\label{fermion-SD}
&& \Sigma(P) = e^2\int dK \,G_{\mu\nu}(Q)\,M_{\mu\tau}\,\gamma_\tau \, S(K) \,\Gamma_\nu\,.
\eea
Comparing (\ref{SF2}, \ref{fermion-SD}) with (\ref{dressedS}) we find the  operators that project out each of the fermion dressing functions. For example
\bea
{\cal P}_{B^+} = \frac{1}{4}(1+\gamma_5) ~~
\to ~~  B_P^+ = {\rm Tr}\big[{\cal P}_{B^+} \Sigma(P)\big]\,.
\eea

Performing the traces we obtain the set of self-consistent integrals that give the six fermion dressing functions:
\bea
\label{Z-sd}
&& Z_P^\pm = 1 - \frac{4\alpha \pi v_F}{2 p_0} \int \frac{dK}{\text{Den}_K{}^\pm}\, \frac{q^2 G_L}{Q^2} k_0  Z_K{}^\pm \left(Z_K{}^\pm + Z_P{}^\pm\right)  \\
&& A_P^\pm = 1 + \frac{4\alpha \pi v_F}{2p^2} \int \frac{dK}{\text{Den}_K{}^\pm}\,\left[ k_0 G_{\text{DE}} (\vec q\times \vec p) Z_K{}^\pm
   \left(A_K{}^\pm +A_P{}^\pm -Z_K{}^\pm - Z_P{}^\pm\right) \right.  \nonumber \\
&& ~~~~~~~~~~\left.   +\frac{G_L}{Q^2 }\left(q^2 (\vec k\cdot\vec p) A_K{}^\pm \left(Z_K{}^\pm + Z_P{}^\pm \right)+k_0 q_0 
(\vec p\cdot\vec q)  Z_K{}^\pm \left(A_K{}^\pm + A_P{}^\pm + Z_K{}^\pm + Z_P{}^\pm \right) \right.\right.\nonumber\\[1mm]
&&   ~~~~  ~~~~~ \left.\left. - q_0 (\vec q\times \vec p) B_K^+ (A_K{}^\pm + A_P{}^\pm - Z_K{}^\pm - Z_P{}^\pm)
\right) \right]\nonumber \\
&& B_P^\pm = \frac{4\alpha \pi v_F}{2} \int \frac{dK}{\text{Den}_K{}^\pm}\,\frac{q^2 G_L}{Q^2} B_K{}^\pm \left(Z_K{}^\pm + Z_P{}^\pm\right)\,.\nonumber
\eea
We have used the notation $\vec q\times\vec p = q_1p_2-q_2p_1$, $G_{DE}=G_D+G_E$, and dropped terms proportional to $v_F^2$ (relative to 1) - which is the reason there are no terms containing factors $G_T$ in (\ref{Z-sd}). 

From equation (\ref{Z-sd}) it is easy to see that if we find a solution for the plus dressing functions $Z^+$, $A^+$ and $B^+$, then we automatically have a solution for the minus dressing functions of the form $Z^- = Z^+$, $A^- = A^+$ and $B^- = -B^+$. We expect therefore that we will always be able to find a chirally symmetric and time reversal violating solution ($B_{\rm even}=0$ and $B_{\rm odd}\ne0$) if we initialize with $Z_{\rm odd}=A_{\rm odd} = B_{\rm even}=0$. We call this solution 1 and write the solutions for the non-zero dressing functions $Z_{\rm even}^{(1)}$, $A_{\rm even}^{(1)}$ and $B_{\rm odd}^{(1)}$. 

We can also see immediately that  a solution with $Z_{\rm odd}=A_{\rm odd} = B_{\rm odd}=0$ should not exist, since setting all odd dressing functions to zero on the right side of equations (\ref{Z-sd}) gives $Z_{\rm odd}(P) = B_{\rm odd}(P)=0$ but
\bea
A_{\rm odd}(P) & =& \frac{4\alpha \pi v_F }{2p^2} \int \frac{dK}{Q^2} \frac{B_{K{\rm even }}}{\text{Den}_{K\rm{even}}}
\bigg[
q_0 G_L (\vec q \times \vec p)
\left(Z_{K{\rm even }} + Z_{P{\rm even} }  -A_{K{\rm even }} - A_{P{\rm even} } \right)   \nonumber\\[2mm]
& + & G_{DE} Q^2 (\vec p \cdot \vec q)
\left(A_{K{\rm even }} + A_{P{\rm even} } + Z_{K{\rm even }} + Z_{P{\rm even} }\right)
\bigg]\,. 
\eea
In the vicinity of the critical point however, where $B_{K{\rm even }}$ is small, we would have $A_{\rm odd}(P)\approx 0$. We therefore expect to get rapid convergence if we start in the vicinity of the critical point and initialize with $Z_{\rm odd}=A_{\rm odd} = B_{\rm odd}=0$. We will call this solution 2. 

We can also show that the two solutions discussed above are approximately the same, except for the reversal of the even/odd parts of the $B$ dressing function. To see this we substitute on the right side of (\ref{Z-sd})
\bea
\label{init-fake}
&&  Z_{\rm odd}^{(2)} = A_{\rm odd}^{(2)} = B_{\rm odd}^{(2)} = 0  \\
&& Z_{\rm even}^{(2)} = Z_{\rm even}^{(1)}\,,~~~A_{\rm even}^{(2)} = A_{\rm even}^{(1)}\,,~~~B_{\rm even}^{(2)} = B_{\rm odd}^{(1)}\, \nonumber
 \eea
which gives
\bea
\label{init-out}
&& Z_{\rm even}^{(2)} = Z_{\rm even}^{(1)}\\
&& B_{\rm even}^{(2)} = B_{\rm odd}^{(1)} \nonumber \\
&& A_{\rm even}^{(2)} = A_{\rm even}^{(1)} + \frac{4\alpha\pi v_F}{2p^2} \int \frac{B^{(1)}_{K{\rm odd}}}{\text{Den}_{K}Q^2} 
\bigg[q_0 G_L (\vec q\times \vec p) \left(A^{(1)}_{K\rm{even}} + A^{(1)}_{P\rm{even}} - Z^{(1)}_{K\rm{even}} - Z^{(1)}_{P\rm{even}}\right) \nonumber \\
&& ~~~~ ~~~-G_{DE} Q^2
   (\vec p\times\vec q) \left(A^{(1)}_{K\rm{even}} + A^{(1)}_{P\rm{even}} + Z^{(1)}_{K\rm{even}} + Z^{(1)}_{P\rm{even}}\right)\bigg] \,. \nonumber
\eea
The first two lines in (\ref{init-out}) are consistent with (\ref{init-fake}), and the last line is approximately consistent when we are close to the critical point. 

This analysis agrees with our numerical results, which are presented in detail in section \ref{results-section}. In summary, for all values of $(\alpha, \theta)$ we have considered, we have only found two solutions which have the form
\bea
\label{soln1}
\hspace*{-1cm} \text{soln 1:~~}&& Z^{(1)}_{\rm even} \ne 0\,,~~ A^{(1)}_{\rm even}\ne 0 \,,~~ B^{(1)}_{\rm odd}\ne  0 \,;~~ Z^{(1)}_{\rm odd}= A^{(1)}_{\rm odd} = B^{(1)}_{\rm even} = 0 \\
\label{soln2}
\hspace*{-1cm} \text{soln 2:~~}&& Z^{(2)}_{\rm even} \approx Z^{(1)}_{\rm even}\,,~~ A^{(2)}_{\rm even} \approx A^{(1)}_{\rm even}\,, ~~ B^{(2)}_{\rm even} \approx  B^{(1)}_{\rm odd}\,; ~~ Z^{(2)}_{\rm odd}\approx A^{(2)}_{\rm odd} \approx B^{(2)}_{\rm odd} \approx 0 \,.
\eea
The symbols  `approximately equal to' in the second line indicate deviations from zero of less than 0.01 percent. Solution 1 preserves chiral symmetry but violates time reversal invariance and, to the degree of accuracy noted above, solution 2 breaks chiral symmetry but satisfies time reversal invariance.

\subsection{Photon Schwinger-Dyson equations}
\label{phot-sd-eqns}

The two components of the polarization tensor denoted $\rho$ and $\Theta$ can be set to zero in the 
approximation $v_F^2<<1$, which is consistent with what was done with the fermion dressing functions in section \ref{ferm-sd-eqns}. In this case equations (\ref{prop}) become
\bea
\label{approx11}
G_L &=& \frac{2 Q + \alpha }{(2 \sqrt{Q^2} + \alpha ) (2 \sqrt{Q^2}+ \alpha +\gamma ) + 4 Q^2 \theta^2} \\
G_D+G_E \equiv G_{DE} &=& -\frac{2 \theta}{(2 \sqrt{Q^2}+ \alpha ) (2 \sqrt{Q^2} + \alpha +\gamma ) + 4 Q^2 \theta^2}\,.
\nonumber
\eea
These expressions involve only two components of the polarization tensor: $\alpha(p_0,p)$ and $\gamma(p_0,p)$. We work with the more convenient expressions 
\bea
&& \Pi_{00} = \frac{q^2}{Q^2}(\alpha+\gamma) \\
&& {\rm Tr} \Pi = \Pi_{\mu\mu} = \alpha +\frac{Q^2}{q^2}\Pi_{00}\,.
\eea
From the Schwinger-Dyson equation for the polarization tensor we obtain
\bea
\label{pi00-sd}
\Pi_{00} && = -4\alpha\pi v_F \\
&& \int \frac{dK }{\text{Den}_K{}^+ \text{Den}_Q{}^+}\,
(Z_K{}^++Z_Q{}^+) \big(v_F^2 (\vec k\cdot \vec q) A_K{}^+ A_Q{}^++B_K{}^+
   B_Q{}^+-k_0 q_0 Z_K{}^+ Z_Q{}^+\big)  + (+\rightarrow -)  \nonumber	
\eea
where the notation $(+\rightarrow -)$ indicates a second integral with the same form as the first but with all plus superscripts changed to minus. Similarly we obtain for the trace
\bea
\label{pitr-sd}
\Pi_{\mu\mu} = -4\alpha\pi v_F && \int \frac{dK}{\text{Den}_K{}^+ \text{Den}_Q{}^+}\,\big[
2 v_F^2 \left(A_K{}^++A_Q{}^+\right) \left(B_K{}^+ B_Q{}^++k_0 q_0 Z_K{}^+
   Z_Q{}^+\right) \\
&& +
\left(Z_K{}^++Z_Q{}^+\right) \big(v_F^2 (\vec k\cdot \vec q) A_K{}^+ A_Q{}^++B_K{}^+
   B_Q{}^+-k_0 q_0 Z_K{}^+ Z_Q{}^+\big)\big]   
+ (+\rightarrow -)\,.\nonumber
\eea
Equations (\ref{Z-sd}, \ref{approx11}, \ref{pi00-sd}, \ref{pitr-sd}) form a complete set of self-consistent equations that involve only the approximation $v_F^2\ll 1$.

Now we discuss some additional approximations for the photon propagator and dressing functions. 
%
From equations (\ref{pi00-sd}, \ref{pitr-sd}) it is straightforward to show that 
\bea
\Pi_{\mu\mu} = \Pi_{00} + {\cal O}(v_F^2)
\eea
and therefore to ${\cal O}(v_F^2)$ we can set $\Pi_{\mu\mu} = \Pi_{00}$ which gives
\bea
\label{alpha-eqn}
\alpha(q_0,q) = -\frac{q_0^2}{q^2}\Pi_{00}\,.\eea
From equation (\ref{alpha-eqn}) we see that by making a coulomb-like approximation we can set $\alpha(q_0,q)=0$. The full coulomb approximation involves setting $q_0=0$ everywhere in the photon propagator. 
We summarize below:
\bea
\begin{array}{|l|c|}
\hline 
\vspace*{-.1cm} & \\
{\rm approx~ 1:~~} (Z,A,B,\Pi)\bigg|_{v_F^2\ll 1}& ~~ G_L = 
\frac{q^2 Q^2 \Pi _{00}-q^4 (\Pi_{\mu\mu}+2 Q)}
{Q \left(Q^3 \Pi_{00}^2-\Pi_{\mu\mu} q^2 \left(Q \Pi _{00}+2 q^2\right)-4 \left(1 + \theta ^2\right) q^4 Q\right)} ~~ \\
 & ~~ G_{DE} =
\frac{2 \theta  q^4}
{Q \left(Q^3 \Pi_{00}^2-\Pi_{\mu\mu} q^2 \left(Q \Pi _{00}+2 q^2\right)-4 \left(1 + \theta ^2\right) q^4 Q\right)} ~~ \\
\vspace*{-.1cm} & \\
\hline 
\vspace*{-.1cm} & \\
{\rm approx~ 2:~~} (Z,A,B,\Pi)\bigg|_{v_F^2\ll 1} ~\Pi_{\mu\mu}=\Pi_{00} & G_L = \frac{q^2 q_0^2 \Pi _{00}-2 q^4 Q}
{Q \left(q_0^2 Q \Pi_{00}^2 -2 q^4 \Pi _{00}-4\left(1 + \theta ^2\right) q^4 Q\right)}  \\
& G_{DE} = \frac{2 \theta  q^4}
{Q \left(q_0^2 Q \Pi_{00}^2 -2 q^4 \Pi _{00}-4\left(1 + \theta ^2\right) q^4 Q\right)}  \\
\vspace*{-.1cm} & \\
\hline 
\vspace*{-.1cm} & \\
{\rm approx ~ 3:~~}  (Z,A,B)\bigg|_{v_F^2\ll 1} 
   \Pi\bigg|_{(v_F^2,q_0/q)\ll 1} & G_L = \frac{q^2}{Q \left(Q \Pi _{00} + 2 \left(1 + \theta ^2\right) q^2 \right)}    \\
& G_{DE} = -\frac{\theta  q^2}{Q^2 \left(Q \Pi _{00} + 2 \left(1+\theta ^2\right) q^2\right)}\\
\vspace*{-.1cm} & \\
\hline 
\vspace*{-.1cm} & \\
{\rm approx~ 4:~~} (Z,A,B,\Pi)\bigg|_{v_F^2\ll 1} ~~(G_L,G_{DE})\bigg|_{q_0=0} & G_L =  \frac{1}{\Pi _{00} + 2(1+\theta^2) q }  \\
 & G_{DE} = -\frac{\theta }{q \left(\Pi _{00} + 2 \left(1 + \theta ^2\right) q \right)}\\
\vspace*{-.01cm} & \\
\hline
\end{array}\nonumber
\label{approxs}
\eea
\normalsize
When $\theta=0$ approximations 1 and 3 reduce to the full back-coupled calculation of Ref. \cite{Carrington1}, and approximation 4 reduces to the Coulomb version of that calculation.

We  also consider using analytic results for the polarization components $\Pi_{00}$ and $\Pi_{\mu\mu}$  obtained from the 1-loop expressions using bare fermion propagators. This is a commonly used approximation, and is based on the vanishing fermion density of states at the Dirac points.

Finally we note that although equation (\ref{approx11}) indicates that $G_{DE}$ is of the same order as $G_L$ for values of $\theta$ of order one, 
we expect that the contribution of this term to the fermion dressing functions will be small. 
To understand this point, recall that the propagator component $G_{T}$ does not contribute in equation (\ref{Z-sd}) because it drops out in the limit $v_F^2<<1$. Likewise  in the second line of (\ref{Z-sd}) the term proportional to $G_{DE}$ is proportional to a difference of the form $Z-A$, and the first two lines of this equation show that this difference is of order $v_F$. 

In summary, the full set of possible approximations we have discussed above can be written using the notation $(n,m,l)$ where:

\noindent {\bf 1)} $n\in(1,2,3,4)$: approximation 1, 2, 3 or 4 [as defined in the table under equation (\ref{alpha-eqn})]. 

\noindent {\bf 2)} $m\in(0,1)$: $m=0$ means the polarization components $\Pi_{00}$ and $\Pi_{\mu\mu}$ are obtained from their self-consistent expressions (back coupled); $m=1$ means we use their analytic 1-loop approximations. 

\noindent {\bf 3)} $l\in(0,1)$: $l=0$ means $G_{DE}$ is set to zero; $l=1$ means $G_{DE}$ is included. 

\noindent There are in principle 16 possible calculations, corresponding to approximations $n=(1,2,3,4)\times m=(0,1)\times l=(0,1)$. 
Approximations $n\in(1,2,3)$ and $l\in(0,1)$ agree to very high accuracy. We show some results for the values of $B^+(0,0)$ which verify this in Table \ref{table-approx}. From this point on we we will consider only approximations $(3,0,0)$, $(4,0,0)$ and $(4,1,0)$. 
\begin{table}[h]
\begin{center}
\begin{tabular}{|c|c|c|}
\hline 
\vspace*{-.2cm} & &  \\ 
~~~~~~approx~~~~~~~ & ~~~~~~~~~~~$(\alpha,\theta) = (4.0,0.2)$ ~~~~~~~~~~~&~~~~~~~~~~~ $(\alpha,\theta)=(3.4,0.6)$ ~~~~~~~~~~~ \\
 \vspace*{-.2cm} & &  \\ 
 \hline 
 \vspace*{-.2cm} & & \\ 
(3,0,0) & 0.00256085   &    0.00036157\\
 \vspace*{-.2cm} & & \\ 
 \hline 
 \vspace*{-.2cm} & & \\ 
(3,0,1) & 0.00255782    &   0.00036102 \\
 \vspace*{-.2cm} & & \\ 
 \hline 
 \vspace*{-.2cm} & & \\ 
(2,0,0) &  0.00256082   &   0.00036167 \\
 \vspace*{-.2cm} & & \\ 
 \hline 
 \vspace*{-.2cm} & & \\ 
(2,0,1) &   0.00255762  &   0.00036080 \\
 \vspace*{-.2cm} & & \\ 
 \hline 
 \vspace*{-.2cm} & & \\ 
(1,0,0) & 0.00256083    &    0.00036168 \\
 \vspace*{-.2cm} & & \\ 
 \hline 
 \vspace*{-.2cm} & & \\ 
(1,0,1) & 0.00255762    &  0.00036080 \\ 
\vspace*{-.02cm} & & \\ 
\hline
\end{tabular}
\end{center}
\caption{Comparison  of $B^+(0,0)$ from different approximations for two different  values of $(\alpha,\theta)$. \label{table-approx}}
\end{table}

\section{Numerical Method}

We need to solve numerically the set of eight coupled equations (\ref{Z-sd}, \ref{pi00-sd}, \ref{pitr-sd}) for the dressing functions $Z^\pm$, $A^\pm$, $B^\pm$, $\Pi_{00}$ and $\Pi_{\mu\mu}$. 
The functions $\Pi_{00}$ and $\Pi_{\mu\mu}$ are renormalized by subtracting the zero momentum value
\bea
&& \Pi^{\rm renorm}_{00}(P) = \Pi_{00}(P)-\Pi_{00}(0) \,,\nonumber\\
&& \Pi^{\rm renorm}_{\mu\mu}(P) = \Pi_{\mu\mu}(P)-\Pi_{\mu\mu}(0) \,.\nonumber
\eea

We work in spherical coordinates and define $\cos(\theta) = \vec p\cdot \vec k/(p k)$ so that the integrals have the form 
\bea
\int dK &&= \frac{1}{(2\pi)^3} \int_{-\infty}^\infty dk_0 \int_0^\infty dk\, k \int_0^{2\pi} d\theta f(k_0,k,\theta) \nonumber\\
&&= \frac{1}{(2\pi)^3} \int_0^\infty dk_0 \int_0^\infty dk \, k\int_0^{2\pi} d\theta \big[ f(k_0,k,\theta) + f(-k_0,k,\theta)\big]\,.
\eea
We use an ultraviolet cutoff $\Lambda$ on all momentum integrals and define dimensionless variables 
$\hat p_0 = p_0/\Lambda$, $\hat p = p/\Lambda$, $\hat k_0 = k_0/\Lambda$ and $\hat k = k/\Lambda$. We also use generically $\hat B = B/\Lambda$, for all components and representations of the mass-like fermion dressing function. 
The hatted momentum and fequency variables range from $10^{-6}$ to one, and to simplify the notation we suppress all hats. 
We use a logarithmic grid in the $k_0$ and $k$ dimensions to increase sensitivity to the infrared. We use Gauss-Legendre integration. 
Dressing functions are interpolated using double linear interpolation, using grids of $220\times200\times16$ points in the $k_0$, $k$ and $\theta$ dimensions. 
In the calculation of $\Pi_{\mu\nu}$ we use an adaptive grid for the $k_0$ integral to more efficiently include the region of the integral where $k_0\sim p_0$
\bea
\int_{10^{-6}}^1 dk_0 = \int_{10^{-6}}^{p_0} dk_0 + \int_{p_0}^1 dk_0\,.
\eea
The integrands for the fermion dressing functions are smoother and the adaptive grid is not needed. 

\section{Results}
\label{results-section}

Unless stated otherwise, all results in this section are obtained with approximation (3,0,0). 

In Refs \cite{Carrington0,Carrington1} we learned that using the Lindhard screening function, instead of calculating the photon polarization tensor using a self-consistently `back-coupled' formulation, produces an artificially large damping effect which increases the critical coupling. This result can be understood as arising from the fact that  large fermion dressing functions $Z$ and $A$ are neglected in the denominator of the integral that gives the Lindhard expression for the polarization tensor. 
In this work we find that higher values of $\theta$ increase the critical coupling, and this result can be understood in the same way as resulting from increased screening. 

\begin{figure}[H]
\begin{center}
\includegraphics[width=18cm]{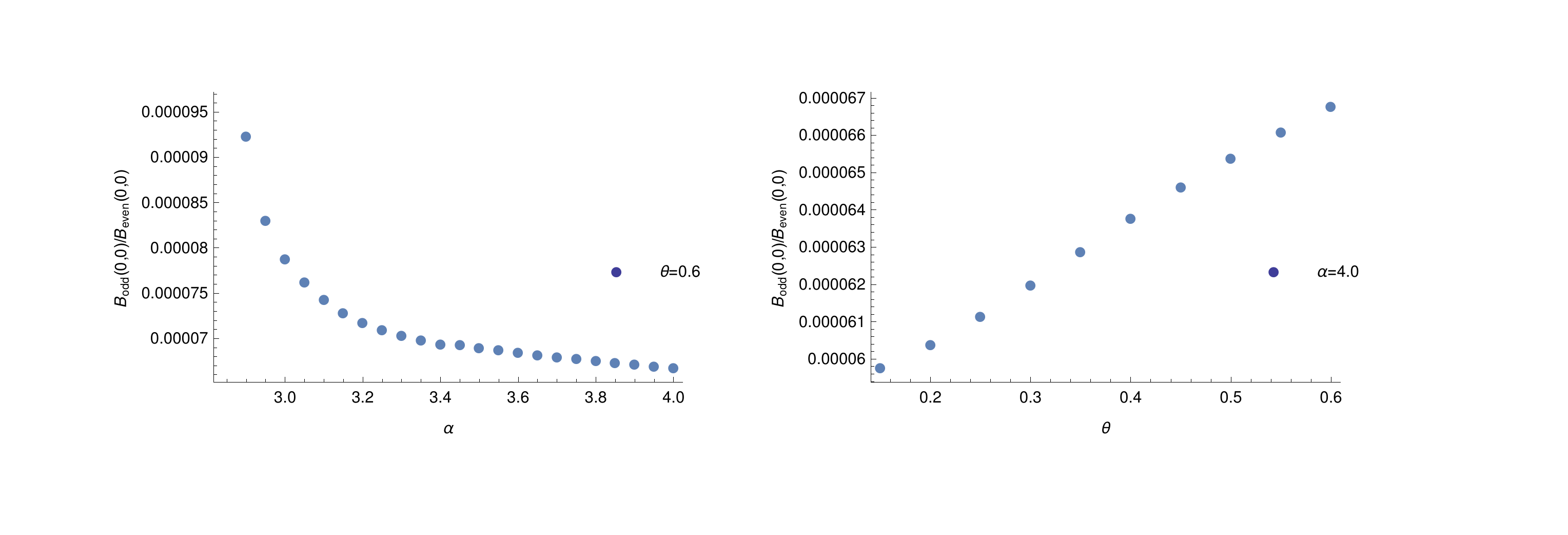}
\end{center}
\caption{The ratio of the odd mass divided by the even one for solution 2 [see equation (\ref{soln2})]. The left panel shows the ratio as a function of the coupling with $\theta=0.6$, and the right panel shows the dependence on $\theta$ with $\alpha=4.0$. \label{evenOodd} }
\end{figure}

We have found (for all values of $\theta$ and $\alpha$ considered) only two types of solutions [see equations (\ref{soln1}, \ref{soln2})]. Up to very small corrections, there is an odd mass solution (solution 1) and an even mass solution (solution 2), but no solutions for which both the even and odd mass parameters are  non-zero. 
In Fig. \ref{evenOodd} we show the absolute value of $B_{\rm odd}(0,0)/B_{\rm even}(0,0)$ for solution 2. As claimed in the text under equation (\ref{soln2}), this ratio is always less than 0.01\%..

\begin{figure}[H]
\begin{center}
\includegraphics[width=10cm]{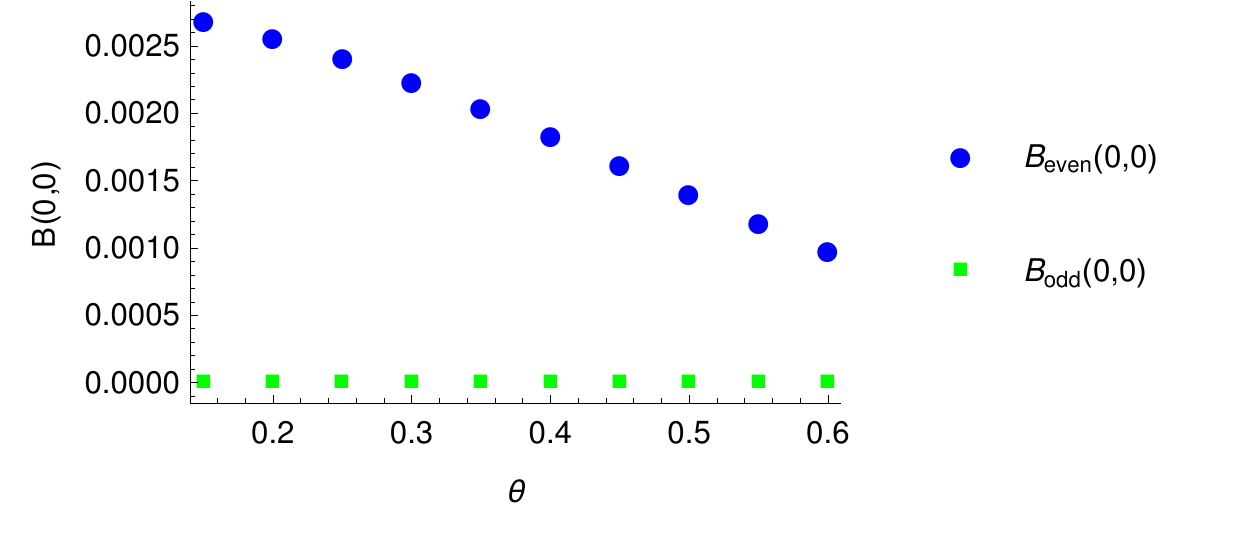}
\end{center}
\caption{$B_{\rm even}(0,0)$ and $B_{\rm odd}(0,0)$ as functions of the parameter $\theta$ with coupling $\alpha = 4.0$. \label{b-th} }
\end{figure}

\begin{figure}[H]
\begin{center}
\includegraphics[width=10cm]{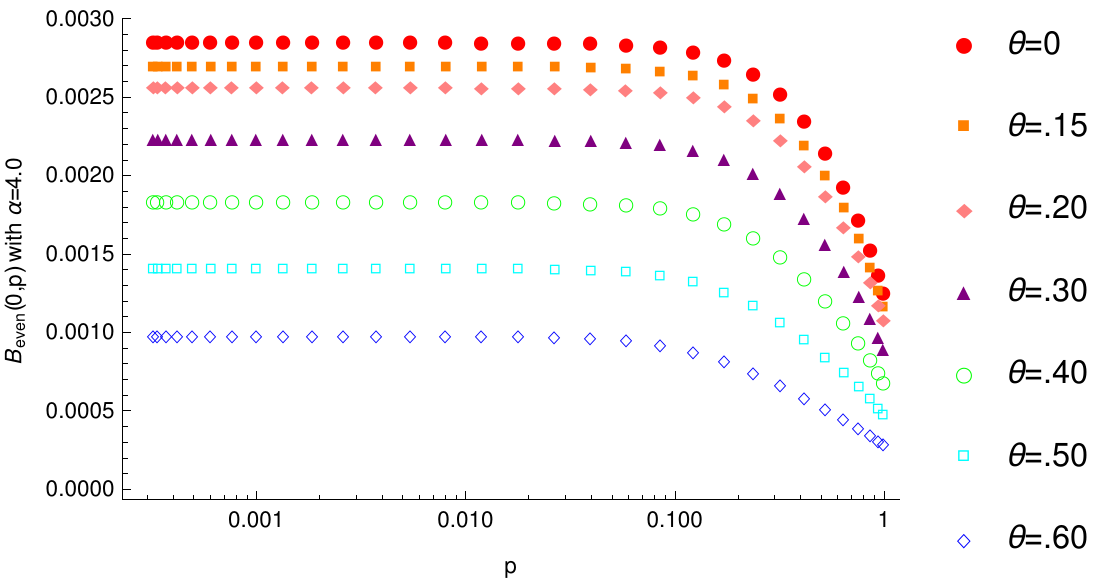}
\end{center}
\caption{$B_{\rm even}(0,p)$ as a function of momentum at fixed $\alpha=4.0$. \label{b-p}}
\end{figure}

From this point on we show only results from solution 2. 
In Fig. \ref{b-th} we show the condensates $B_{\rm even}(0,0)$ and $B_{\rm odd}(0,0)$ as a function of $\theta$ at fixed coupling, and in Fig. \ref{b-p} we show the dressing function $B_{\rm even}(p_0,p)$ as a function of momentum at fixed $p_0=0$, using different values of $\theta$. These figures show clearly that the condensate decreases as a function of $\theta$, which implies that the critical coupling will increase as $\theta$ increases. 

\begin{figure}[H]
\begin{center}
\includegraphics[width=12cm]{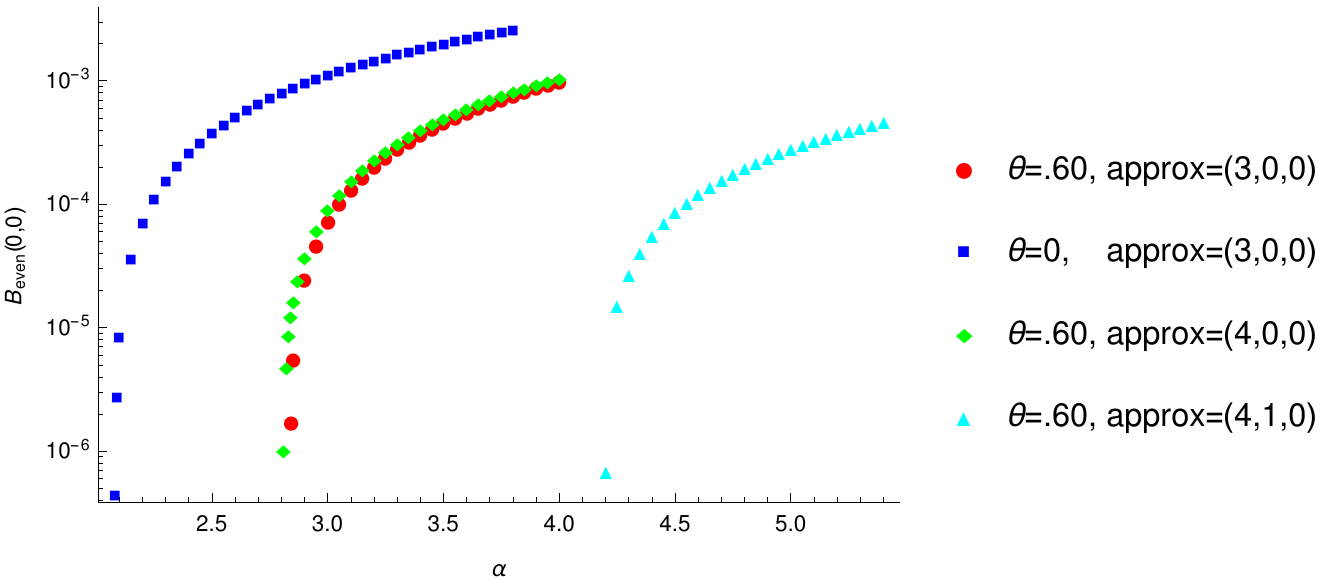}
\end{center}
\caption{$B_{\rm even}(0,0)$ as a function of coupling for different approximations and different values of the parameter $\theta$.  \label{b-al}}
\end{figure}

The dependence of the critical coupling on the parameter $\theta$ is seen explicitly in Fig. \ref{b-al} which shows the condensate as a function of $\alpha$ for different values of  $\theta$, using different approximations.

\begin{figure}[H]
\begin{center}
\includegraphics[width=10cm]{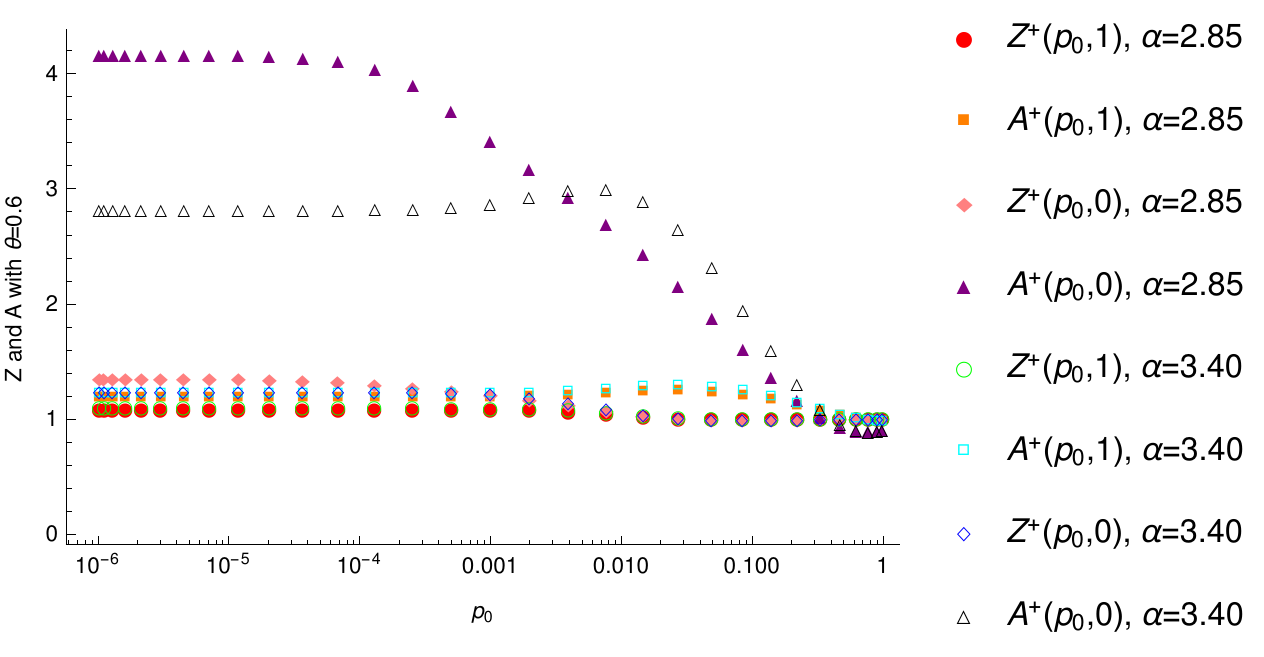}
\end{center}
\caption{The dressing functions $Z^+$ and $Z^-$ as functions of $p_0$, with $p$ held fixed to its maximum and minimum values, for two values of $\alpha$ and $\theta=0.6$.  \label{ZA-p0}}
\end{figure}

\begin{figure}[H]
\begin{center}
\includegraphics[width=10cm]{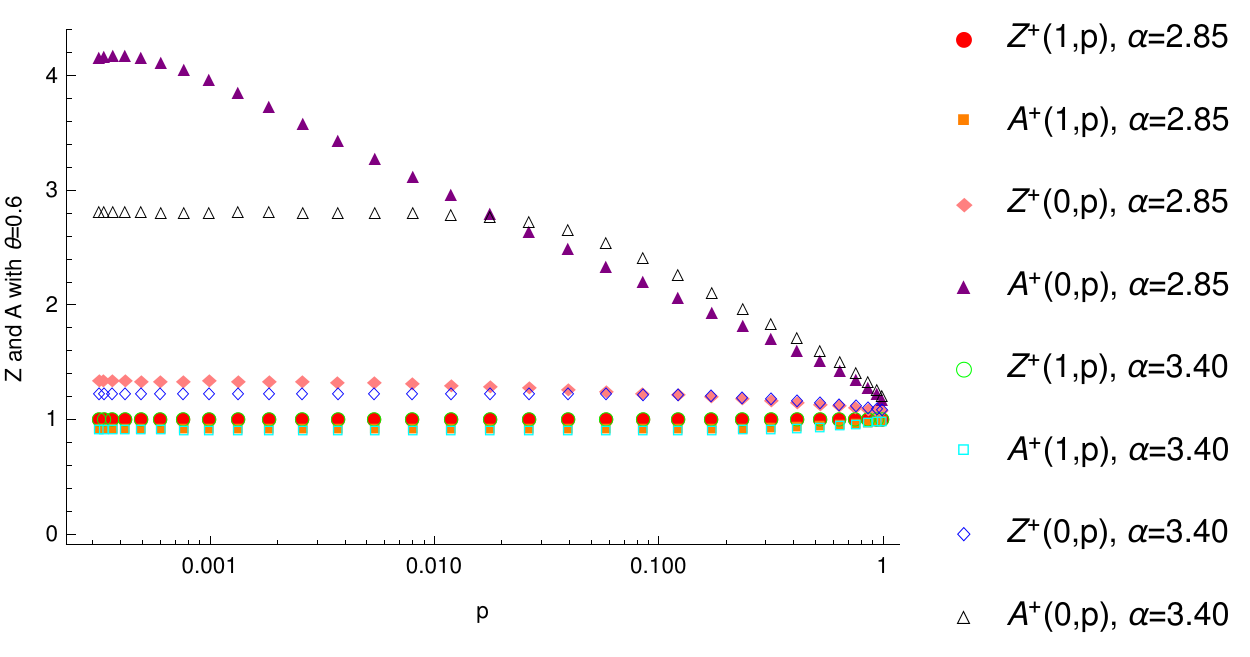}
\end{center}
\caption{The dressing functions $Z^+$ and $A^+$ as functions of $p$, with $p_0$ held fixed to its maximum and minimum values, for two values of $\alpha$ and $\theta=0.6$.  \label{ZA-p}}
\end{figure}

In order to understand what drives this behaviour, we look at the momentum dependence of the dressing functions $Z$ and $A$. Figs. \ref{ZA-p0} and \ref{ZA-p} show the dressing functions $Z^+$ and $A^+$ as functions of $p_0$ and $p$ with the other variable held fixed to its maximum or minimum value. The two values of $\alpha$ that are shown are $\alpha=2.85$, which is close to the critical coupling for the value of $\theta=0.6$ that is chosen, and $\alpha=3.4$ which is relatively far from the critical coupling. One sees that the $Z$ dressing function does not change much, but the $A$ function does change and is responsible for the experimentally observed increase in the Fermi velocity at
small frequencies as one approaches the critical coupling. 

\begin{figure}[H]
\begin{center}
\includegraphics[width=10cm]{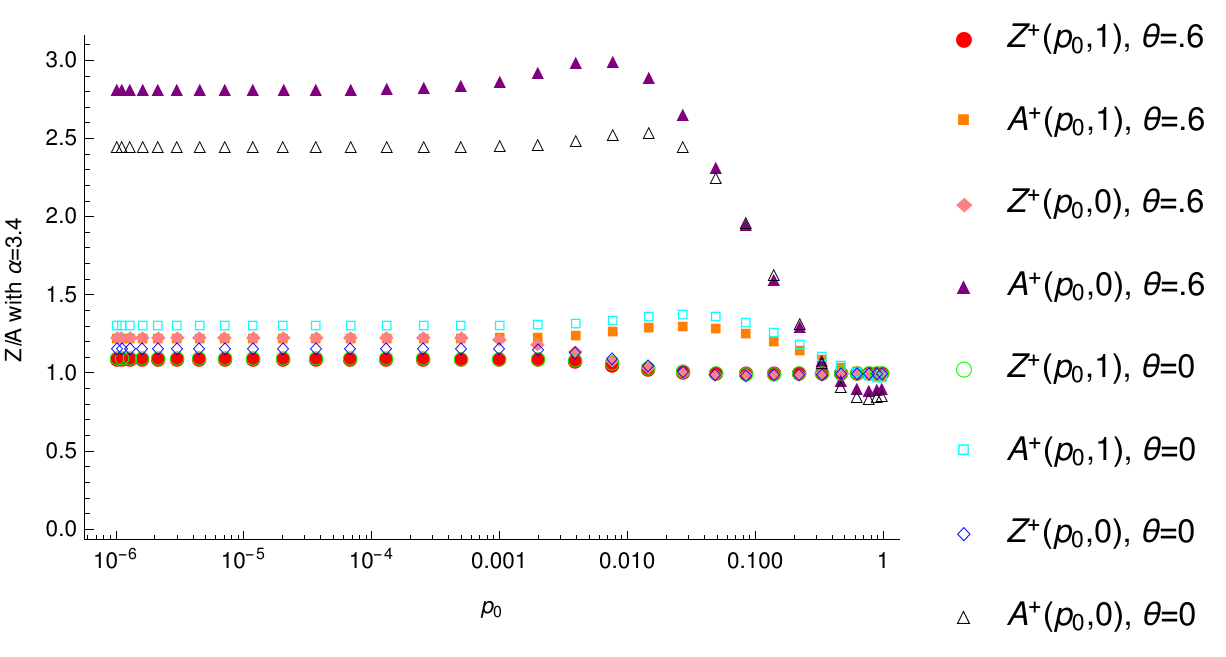}
\end{center}
\caption{The dressing functions $Z^+$ and $A^+$ as functions of $p_0$ for $\alpha=3.4$ and $\theta\in (0,0.6)$.  \label{ZA-th}}
\end{figure}

To see explicitly how this effect is influenced by the parameter $\theta$, we show in Fig. \ref{ZA-th} the fermion dressing functions $Z^+$ and $A^+$ as functions of $p_0$ for two different values of $\theta$. The figure shows that once again it is  $A^+(p_0,0)$ which changes the most, and that the largest effect is obtained with the higher value of $\theta$. 

\begin{figure}[H]
\begin{center}
\includegraphics[width=15cm]{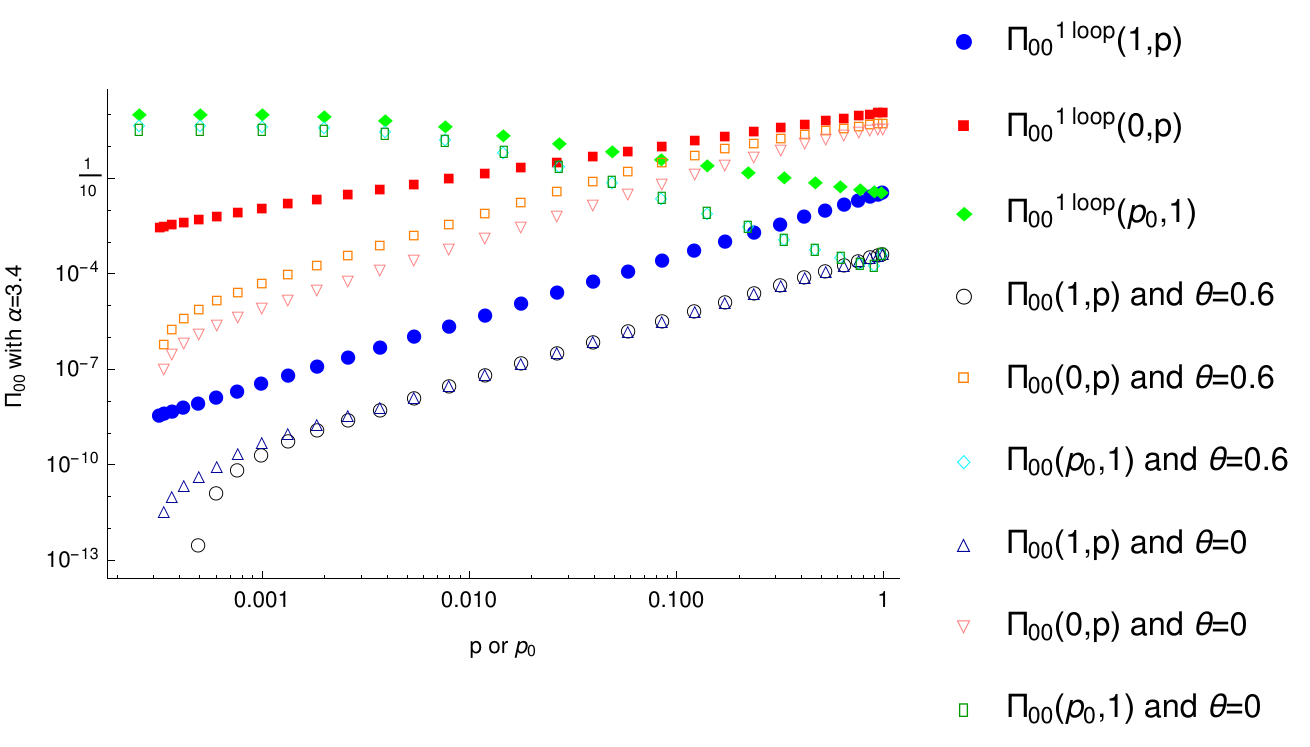}
\end{center}
\caption{The component $\Pi_{00}$ as a function of $p_0$ and $p$ for $\alpha=3.4$ and $\theta\in (0,0.6)$.  \label{pi-all}}
\end{figure}

In Fig. \ref{pi-all} we show $\Pi_{00}$ as a function of momentum for $\alpha=3.4$ and two different values of $\theta$. For comparison the Lindhard expression is also shown. Maximal screening is obtained with the Lindhard approximation, and the smallest screening effect occurs when we set $\theta$ to zero. This is consistent with our results in Figs. \ref{b-th} and \ref{b-al} which show that the critical coupling increases with $\theta$.

We fit the data shown in Fig. \ref{b-al} using Mathematica and the resulting function is extrapolated to obtain the value of the critical coupling for which $B_{\rm even}(0,0)$ goes to zero. Our results are collected in Table \ref{table-alphaC}. The result for approximation (4,1,0) with $\theta=0$ is taken from \cite{Carrington0} and the result for approximation (4,0,0) with $\theta=0$ is taken from \cite{Carrington1}.
\begin{table}[h]
\begin{center}
\begin{tabular}{|c|c|c|c|}
\hline 
\vspace*{-.1cm} & & & \\ 
\diagbox{~~~$\theta$}{~~approx~~} & ~~~~~(3,0,0) ~~~~~&~~~~~ (4,0,0) ~~~~~&~~~~~ (4,1,0)~~~~~ \\ 
[-.2em] & & & \\ \hline
\vspace*{-.1cm} & & & \\ 
$0.0$ & 2.07 & 1.99 & 3.19 \\ 
[-.1em] & & & \\ \hline
\vspace*{-.1cm} & & & \\ 
$0.6$ & 2.84 & 2.80  & 4.20 \\ 
[-.1em] & & & \\ \hline
\end{tabular}
\end{center}
\caption{Extrapolated values of the critical coupling for different approximations and different values of the Chern-Simons parameter. \label{table-alphaC}}
\end{table}

\section{Conclusions}

Chern-Simons terms have been widely studied in condensed matter physics in the context of chiral symmetry breaking, the Hall effect and high temperature superconductivity. In this work we show for the first time that they are also relevant to the study of phase transitions in graphene. 
We work with a low energy effective theory that describes some features of mono-layer suspended graphene. We use reduced QED$_{3+1}$ which describes planar electrons interacting with photons that can propagate in three spatial dimensions. 
We have studied the effect of a Chern-Simons term in this theory. 
We have found two classes of solutions: in the odd sector the theory dynamically generates a time-reversal violating Haldane type mass, and in the even sector a mass term of the standard Dirac type is generated. We have studied the dependence of the Dirac mass on the Chern-Simons parameter ($\theta$) and shown that it is suppressed as $\theta$ increases, which means that the critical coupling at which a non-zero Dirac condensate is generated increases with $\theta$. We have shown that this effect can be understood physically as arising from an increase in screening. \\

{\bf Acknowledgements:}
This work has been supported by the Natural Sciences and Engineering Research Council of Canada and the Helmholtz International Center for FAIR.
The author thanks C.S. Fischer, L. von Smekal and M.H. Thoma for hospitality at the Institut f\"{u}r Theoretische Physik, Justus-Liebig-Universit\"{a}t Giessen, and for discussions. 

\appendix

\section{Notation}
\label{notation}

Our definitions of the lattice vectors are:
\bea
&& a_1 = a \left\{-\sqrt{3} ,0,0\right\}\\
&& a_2 = \frac{a}{2} \left\{-\sqrt{3},-3,0\right\} \nonumber \\
&& a_3 = \frac{a}{2} \left\{3 \sqrt{3},3,0\right\}\,.\nonumber
\eea
Using these definitions the volume of the lattice cell is $S=\frac{3 \sqrt{3} a^2}{2}$.
The vectors that generate the positions of the nearest neighbour lattice points are
\bea
&& \delta_1 = \frac{a}{2} \left\{-\sqrt{3},1,0\right\} \\
&& \delta_2 = \{0,-a,0\} \nonumber \\
&& \delta_3 = \left\{\sqrt{3},1,0\right\} \nonumber
\eea
and the reciprocal lattice vectors are:
\bea
&& b_1 = \frac{2\pi}{a} \left\{-\frac{1}{\sqrt{3}},\frac{1}{3},0\right\} \\
&& b_2 = \frac{2\pi}{a} \left\{0,-\frac{2}{3},0\right\} \nonumber \\
&& b_3 = \frac{2\pi}{a} \left\{-\frac{1}{\sqrt{3}},-\frac{1}{3},0\right\}\,. \nonumber
\eea

The six $K$ points are:
\bea
K_i =  \frac{3a}{2\pi} 
\left(
\begin{array}{cc}
 \frac{1}{\sqrt{3}} & 1 \\
 \frac{4}{\sqrt{3}} & 0 \\
 \frac{1}{\sqrt{3}} & -1 \\
 -\frac{1}{\sqrt{3}} & -1 \\
 -\frac{4}{\sqrt{3}} & 0 \\
 -\frac{1}{\sqrt{3}} & 1 \\
\end{array}
\right)\,
\eea
and we choose our two inequivalent $K$ points as 
\bea
\label{Kpoints}
K_+ = -K_- = \left\{-\frac{8 \pi }{3 \sqrt{3} a},0\right\}\,.
\eea

We define the Fourier transform 
\bea
\label{fourier1}
&& a_{\vec n \sigma} = \sqrt{S}\int_{\rm BZ} \frac{d^2 k}{(2\pi)^2} e^{i\vec k\cdot\vec n} a_\sigma(\vec k) \\
\label{fourier2}
&& \sum_{\vec n} e^{i(\vec k-\vec k')\cdot\vec n} = \frac{(2\pi)^2}{S} \delta^2(\vec k - \vec k')\,. \nonumber
\eea

Our representation of the $\gamma$ matrices is
\bea
&& \gamma^0 = \left(
\begin{array}{cccc}
 0 & 0 & 1 & 0 \\
 0 & 0 & 0 & 1 \\
 1 & 0 & 0 & 0 \\
 0 & 1 & 0 & 0 \\
\end{array}
\right),
\gamma^1 = \left(
\begin{array}{cccc}
 0 & 0 & 0 & -1 \\
 0 & 0 & -1 & 0 \\
 0 & 1 & 0 & 0 \\
 1 & 0 & 0 & 0 \\
\end{array}
\right),
\gamma^2 = \left(
\begin{array}{cccc}
 0 & 0 & 0 & i \\
 0 & 0 & -i & 0 \\
 0 & -i & 0 & 0 \\
 i & 0 & 0 & 0 \\
\end{array}
\right),                 \nonumber      \\
&& \gamma^3 = \left(
\begin{array}{cccc}
 0 & 0 & -1 & 0 \\
 0 & 0 & 0 & 1 \\
 1 & 0 & 0 & 0 \\
 0 & -1 & 0 & 0 \\
\end{array}
\right),
\gamma^5 = \left(
\begin{array}{cccc}
 1 & 0 & 0 & 0 \\
 0 & 1 & 0 & 0 \\
 0 & 0 & -1 & 0 \\
 0 & 0 & 0 & -1 \\
\end{array}
\right)\,.
\eea

\section{Haldane type mass}
\label{haldane-mass}

We consider a term in the Hamiltonian which would give counter-clockwise hopping around the triangles that are formed by each sublattice. We write
\bea
H_2 && = t_2\sum\big[i (a^\dagger_{x_1}a_{x_2} + a^\dagger_{x_2}a_{x_3} + a^\dagger_{x_3}a_{x_1})\big]  + t_2\sum\big[i (b^\dagger_{y_1}b_{y_2} + b^\dagger_{y_2}b_{y_3} + b^\dagger_{y_3}b_{y_1})\big]~+~{\rm h.c.} \nonumber
\label{discrete-haldane}\\
\eea
where $\{\vec x_1,\vec x_2,\vec x_3\}$ and $\{\vec y_1,\vec y_2,\vec y_3\}$ indicate the $A$ and $B$ sites on one hexagonal cell, and the sums are over all $A$ and $B$ triangular sublattices. 
We will take the origin of the coordinate system to be at $\vec x_1 = (0,0)$.
Using our definitions of the lattice vectors, the corners of the trianglar $A$ and $B$ sublattices which form the hexagon with $\vec x_1$ at the lower left corner are
\bea
&& \vec x_1 = (0,0) \\
&& \vec x_2 = -\frac{\vec a_1}{\sqrt{3}} = a(1,0) \nonumber\\
&& \vec x_3 = -\frac{\vec a_2}{\sqrt{3}} = \frac{a}{2}(1,\sqrt{3}) \nonumber\\
&& \vec y_1 = -\frac{2}{3}\vec a_1 + \frac{1}{3}\vec a_2 = \frac{a}{2\sqrt{3}}(\sqrt{3},-1) \nonumber\\
&& \vec y_2 = -\frac{2}{3}\vec a_1 - \frac{2}{3}\vec a_2 = \frac{a}{\sqrt{3}}(\sqrt{3},1) \nonumber\\
&& \vec y_3 = -\frac{1}{3}\vec a_1 - \frac{2}{3}\vec a_2 = \frac{a}{\sqrt{3}}(0,1)\,. \nonumber
\eea
Fourier transforming to momentum space and expanding around the Dirac points we obtain
\bea
H_2 &&  = t_2 {\cal C}  \int \frac{d^2p}{(2\pi)^2} \big[\big(a^\dagger_+(p)  a_+(p) - a^\dagger_-(p)  a_-(p)\big) - \big(b^\dagger_+(p)  b_+(p) - b^\dagger_-(p)  b_-(p)\big)\big] \nonumber\\
&&  =  t_2 {\cal C} \int \frac{d^2p}{(2\pi)^2} \big[\bar\Psi(p) \gamma^3\gamma^5 \Psi(p)\big] \,.
\label{haldane-out}
\eea
where we have defined the constant ${\cal C} = 2(\sin(2\phi)-2\sin(\phi))$; $\phi = 8\pi/(3\sqrt{3})$.
Equation (\ref{haldane-out}) shows that the Hamiltonian (\ref{discrete-haldane}) corresponds to a mass of the form ${\cal M}^{35}$ in the effective theory.


\begin{thebibliography}{9}

\bibitem{pisarski} R.D. Pisarski, Phys. Rev {\bf D29}, 2423 (1984).
\bibitem{appelquist1} T.W. Appelquist, M. Bowick, D. Karabali and L.C.R. Wijewardhana, Phys. Rev. {\bf D33}, 3704 (1986).
\bibitem{fradkin} E. Fradkin and A. L\'{o}pez, Phys. Rev. {\bf B44}, 5246 (1991).
\bibitem{gusynin}  V. P. Gusynin, P. K. Pyatkovskiy,  Phys. Rev. {\bf D94}, 125009 (2016).
\bibitem{teber} A. V. Kotikov and S. Teber,  Phys. Rev. {\bf D94}, 114011 (2016).

\bibitem{appelquist2} T. Appelquist, M.J. Bowick, D. Karabali and L.C.R. Wijewardhana, Phys. Rev. {\bf D33}, 3774 (1986).
\bibitem{matsuyama} T. Matsuyama and H. Nagahiro, Mod. Phys. Lett. {\bf A15}; Grav. Cosmol. {\bf 6}, 145 (2000).
\bibitem{bashir} A. Bashir, A. Raya, S. S\'{a}nchez-Madrigal, J. Phys. {\bf 41}, 505401 (2008).
\bibitem{kondo} K.I. Kondo and P. Maris, Phys Rev. Lett. {\bf 74}, 18 (1995); Phys. Rev. {\bf D52}, 1212 (1995).

\bibitem{highTc} A.P. Balachandran, 
E. Ercolessi, G. Morandi and 
A.M. Srivastava, Int. J. Mod. Phys {\bf B4}, 2057 (1990).
\bibitem{Witten-review} E. Witten, La Rivista del Nuovo Cimento, 39, 313 (2016). 

\bibitem{hosotani} Y. Hosotani, Phys. Lett. {\bf B319}, 332 (1993).
\bibitem{shovkovy}  I.A. Shovkovy, Lect. Notes Phys. 871, 13 (2013).

\bibitem{carbotte} V.P. Gusynin, S.G. Sharapov, and J.P. Carabotte. Int. J. Mod. Phys. {\bf B21}, 4611 (2007).

\bibitem{haldane86} F.D.M. Haldane, Phys. Rev. Lett. {\bf 61}, 2015 (1988). 

\bibitem{marino} E.C. Marino, Nucl. Phys. {\bf B408}, 551 (1993).
\bibitem{Miransky2001} E.V. Gorbar, V.P. Gusynin, V.A. Miransky and I.A. Shovkovy, Phys. Rev. {\bf B66}, 045108 (2002).

\bibitem{ball-chiu}  J.S. Ball and T.W Chiu, Phys Rev. {\bf D22}, 2542 (1980); Phys Rev. {\bf D22}, 2550 (1980).

\bibitem{Carrington0}  M.E. Carrington, C.S. Fischer, L. von Smekal, M.H. Thoma, Phys. Rev. {\bf B94}, 125102 (2016). 

\bibitem{Carrington1} M.E. Carrington, C.S. Fischer, L. von Smekal, M.H. Thoma, Phys. Rev. {\bf B97}, 115411 (2018). 





\end{thebibliography}
\end{document}